\documentclass[aps,pra,twocolumn,amsmath,amssymb,superscriptaddress,floatfix]{revtex4-1}

\usepackage{graphicx}
\usepackage{multirow}
\usepackage{color}
\usepackage{bm}

\renewcommand{\Vec}[1]{\mbox{\boldmath$#1$}}
\def\infinity{\infty}
\def\t#1{\textrm{#1}}
\def\ket#1{|#1\rangle }
\def\bra#1{\langle #1 |}
\def\braket#1{\langle #1 \rangle}
\def\n{\nonumber \\ }

\begin{document}

\title{
Weyl Mott Insulator
}

\author{Takahiro Morimoto}
\affiliation{RIKEN Center for Emergent Matter Science 
(CEMS), Wako, Saitama, 351-0198, Japan}
%\author{Bohm-Jung Yang}
%\affiliation{RIKEN Center for Emergent Matter Science 
%(CEMS), Wako, Saitama, 351-0198, Japan}
\author{Naoto Nagaosa}
\affiliation{RIKEN Center for Emergent Matter Science 
(CEMS), Wako, Saitama, 351-0198, Japan}
\affiliation{Department of Applied Physics, The University of 
Tokyo, Tokyo, 113-8656, Japan}

\date{\today}

\begin{abstract}
Relativistic Weyl fermion (WF) often appears in the band structure of 
three dimensional magnetic materials and acts as a source or sink of the 
Berry curvature, i.e., the (anti-)monopole.
It has been believed that the WFs are stable due to their  
topological indices except when two Weyl fermions of opposite   
chiralities annihilate pairwise.  
Here, we theoretically show for a model including the electron-electron 
interaction that the Mott gap opens for each WF without violating the
topological stability, leading to a topological Mott insulator dubbed 
{\it Weyl Mott insulator } (WMI). This WMI is characterized by several
novel features such as (i) energy gaps in the angle-resolved photo-emission
spectroscopy (ARPES) and the optical conductivity, 
(ii) the nonvanishing Hall conductance, 
and (iii) the Fermi arc on the surface with the penetration depth diverging as 
approaching to the momentum at which the Weyl point is projected.
Experimental detection of the WMI by distinguishing from conventional 
Mott insulators is discussed with possible relevance to pyrochlore iridates.
\end{abstract}

% insert suggested PACS numbers in braces on next line
%72.10.-d 	Theory of electronic transport; scattering mechanisms
%73.20.-r 	Electron states at surfaces and interfaces
%73.43.Cd   Theory and modeling (QHE)
\maketitle

%\section{Introduction}
\textit{Introduction ---}
Weyl fermions (WFs) in solids attract recent intensive interests from the 
viewpoint of their novel quantum transport properties and chiral anomaly. 
The WF is described by the 2-component spinors originating from 4-component 
Dirac spinor 
when the mass $m$ is zero in the Dirac equation. 
%Its Hamiltonian $H$ is given by 2 $\times$ 2 Pauli matrices 
%$\sigma^i $ $(i = 1,2,3)$ as $H = \sum_i \sigma^i p_i$ 
%where $p_i = - i \hbar \partial_i $
%is the momentum operator. 
%It has been proposed that the WF can be 
%a model for neutrino but the confirmation of the WF still remains an 
%unsettled issue in high energy physics. 
The realization of WFs in condensed matters has been recently 
established~\cite{FangScience,Murakami,Wan11}.
In magnetic materials, the time-reversal symmetry is broken and the
energy dispersion of Bloch wavefunction has no Kramer's degeneracy. 
In this case, the band crossings between the two bands are described by
a 2 $\times$ 2 Hamiltonian as 
\begin{equation}
 H(\bm{k})  = \varepsilon_0(\bm{k}) + \sum_{i=1,2,3} \sigma^i h_i(\bm{k}),
\label{eq:bandcross}
\end{equation}
with 2 $\times$ 2 Pauli matrices $\sigma^i $ $(i = 1,2,3)$.
Three conditions of the band crossing $h_i (\bm{k} ) = 0$ for $(i = 1,2,3)$ 
can be satisfied in general by appropriately choosing the three components 
of the crystal momentum $\bm{k}$.
%the codimension 3 matches the dimensionality of the space.
Weyl points sometimes exist 
exactly at the Fermi energy when dictated by some symmetry and topology of the 
Bloch wavefunctions, for example, in
Dirac semimetals~\cite{Young,Liu,Xu-Na3Bi,Neupane,Borisenko}.
More recently, experimental discovery of Weyl semimetals 
in an inversion broken material TaAs has been reported~\cite{TaAs1,TaAs2,Ding}. 
    
Weyl fermion plays an important role in the context of the Berry phase, which is defined 
by $\bm{a}_{n \bm{k}} = - i \bra{ u_{n \bm{k}} } 
\nabla_{\bm{k}} \ket{ u_{n \bm{k}} }$
($ \ket{ u_{n \bm{k}} } $ : the periodic part of the Bloch wave function 
with the band index $n=\pm$ and the momentum $\Vec k$ )
and acts as the vector potential in the momentum space.
The Berry curvature $\bm{b}_{n \bm{k}} = 
\nabla_{\bm{k}} \times \bm{a}_{n \bm{k}}$
is the emergent magnetic field, and can be enhanced near the band crossing points.
When one expand Eq.~(\ref{eq:bandcross}) around the band crossing point 
(Weyl point, which we assume to be $\bm{k}_0 =\bm{0}$), 
there appears the WF described with  
$h_i(\bm{k})=\eta v_F k_i$,
by an appropriate choice of the coordinate $k_i$'s,
where $v_F$ is the Fermi velocity.
The sign $\eta =\pm 1$ specifies the chirality of the WF,
and the Berry curvature of the lower eigenstates $(n = -)$
of the Hamiltonian in Eq.~(\ref{eq:bandcross}) is obtained as
\begin{equation}
 \bm{b}_{- \bm{k}}  = \frac{\eta}{2} \frac{\bm{k}}{| \bm{k}|^3} ,
\label{eq:monopole}
\end{equation}
which diverges as $|\bm{k}| \to 0$ and the total flux $\Phi$
penetrating the surface $S$ enclosing the Weyl point is given by 
$\Phi = \int_S d \bm{S} \cdot \bm{b}_{- \bm{k}} = 2 \pi \eta$.
This indicates that the WF acts as the magnetic monopole (anti-monopole)
for $\eta = 1$ ($\eta = -1$);
the magnetic charge $n_{m} = \frac{\Phi}{2 \pi}$ 
plays a role of topological index. 
Strong Berry curvature leads to the enhanced anomalous 
Hall effect~\cite{FangScience}
as well as the chiral magnetic effect which results in the negative 
magneto-resistance~\cite{Burkov}.  
 
Figure 1 shows the schematic figure of the three dimensional 
first Brillouin zone in which two WFs exist along the $k_z$ direction.
One can define the Chern number 
\begin{equation}
Ch(k_z) = \int \frac{d k_x d k_y}{ 2 \pi} b_z (\bm{k}) ,
\end{equation}
for the plane of fixed $k_z$. When we consider $Ch(k_z)$ 
as a function of $k_z$, there appears the jump by $\eta$ at 
$k_z = \pm k_{0 z}$, i.e., $k_z$-coordinate of the Weyl points. 
Therefore, due to the periodicity of $Ch(k_z)$ by $k_z\to k_z+2 \pi/c$,
we need the pair of $\eta=1$ and $-1$~\cite{Nielsen1,Nielsen2}.
The existence of a single
(an odd number of) WF is also excluded.
Therefore, the annihilation of a single WF is prohibited, i.e.,
the only way to destroy the WFs is to annihilate a
pair of WFs with opposite chiralities either by making the two
WFs approach to each other in the momentum space or 
by introducing a scattering between two WFs with some density-wave-type order.  
The former scenario is actually proposed for the transition between the
Weyl metal and insulator in pyrochlore compounds~\cite{WWK12}.
The latter one is also discussed intensively~\cite{Maciejko,Sekine}.
Meanwhile, effects of the electron correlation have been discussed for WFs by several methods including random phase approximation~\cite{Abrikosov71} and, more recently, cluster perturbation theory~\cite{Witczak-Krempa14}.
However, the possibility of the Mott gap opening at each 
WF has never been explored thus far to the best of the present 
authors' knowledge.

In this paper, we study the effect of the electron correlation $U$ on WFs 
by using a simple model which is exactly solvable. 
It is shown that the Mott gap due to $U$ open at each WF without the
pair annihilation, while the topological properties are kept unchanged.
Namely, the magnetic charge of the WF is unchanged with the role of poles
in Green's function being replaced by its zeros.  
The Hall conductance $\sigma_{xy}$ remains nonvanishing
and the Fermi arc on the surface remains,
while the Green's function and the optical 
conductivity $\sigma_{xx}(\omega)$ show the gap.
%Chiral anomaly of WFs continue to exist. 
Therefore, this Mott insulating state is identified as a topological Mott insulator,
and we name it ``Weyl Mott Insulator (WMI)''.
The experimental detection of this 
new state is also discussed.

\begin{figure}[tb]
\begin{center}
\includegraphics[width=0.8\linewidth]{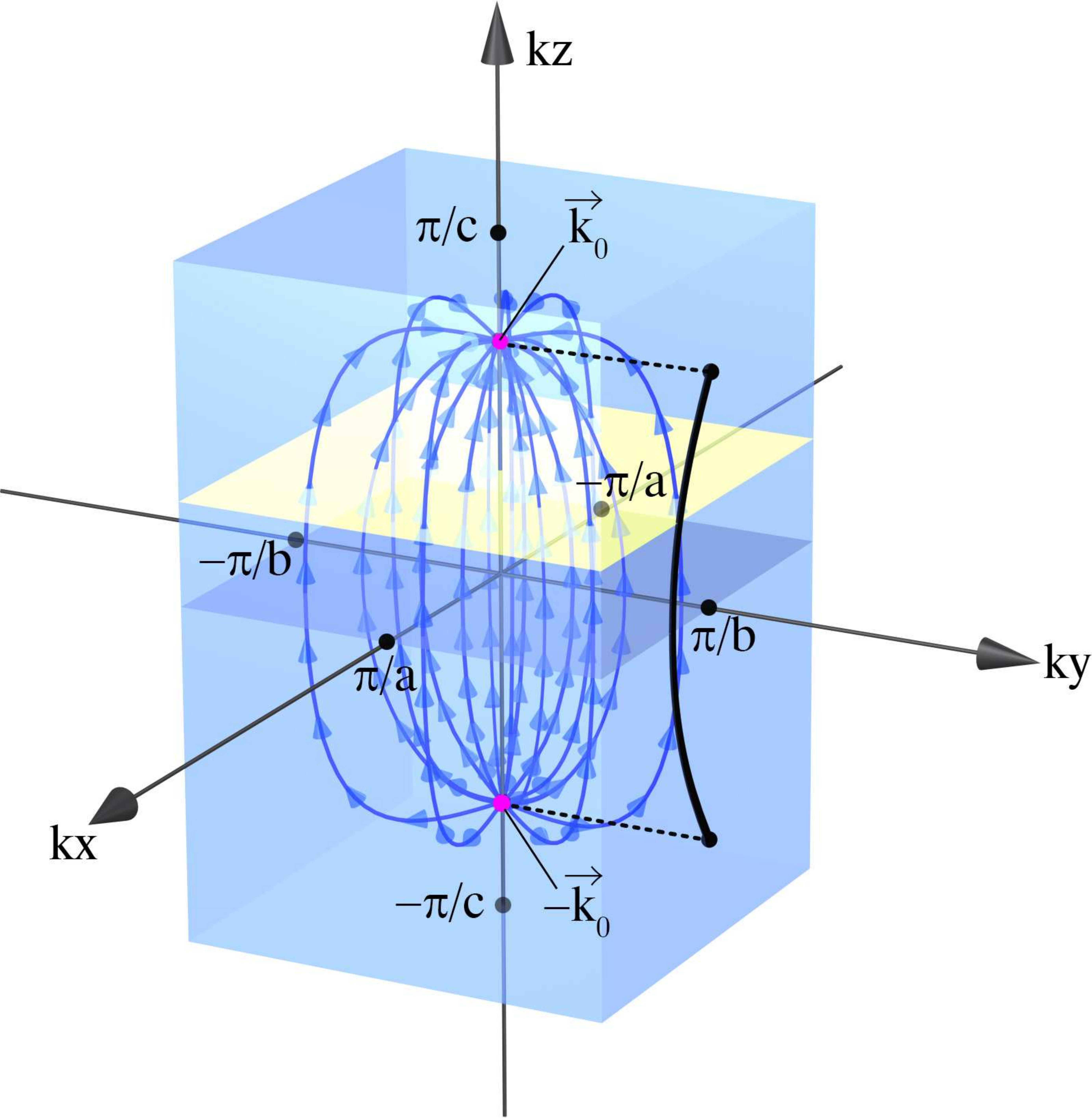}
\end{center}
\caption{
Schematic picture for the Weyl semimetal.
A Weyl point plays a role of a source or sink of the Berry curvature, 
i.e., the (anti-)monopole in the momentum space. 
A pair of Weyl points with opposite charges is accompanied 
with a Fermi arc (the curve on the right side surface). 
}
\label{Fig: schematics}
\end{figure}

\textit{Model and Green's function ---}
The model we study is given by the Hamiltonian
\begin{equation}
H = \sum_{\bm{k}} \left[ 
\psi^\dagger_{\bm{k}} \bm{h}(\bm{k}) \cdot
\bm{\sigma} \psi + \frac{1}{2} U ( n_{\bm{k}} - 1)^2
\right],
\label{eq:Hamiltonian}
\end{equation}
where $\psi_{\bm{k}} = ( c_{\bm{k}, \uparrow},  c_{\bm{k}, \downarrow})^T$
is the two-component spinor, and 
$n_{\bm{k}} = \psi^\dagger_{\bm{k}} \psi^{\,}_{\bm{k}}
=  n_{\bm{k}, \uparrow} +  n_{\bm{k}, \downarrow}$.
The most peculiar nature of this model arises from
the electron-electron interaction 
which is local in $\bm{k}$, i.e., the Hamiltonian is decomposed into
independent $\bm{k}$-sectors. 
In the real space, this corresponds to the
non-local interaction in the limit of forward scattering.
A similar idea has been explored to study the Mott transition~\cite{Hatsugai}
and the spin-charge separation~\cite{Baskaran}.
This locality of the interaction in $\bm k$
enables the exact solution of this problem.
One can introduce the unitary transformation 
$U(\bm{k})$ satisfying $ U(\bm{k})^\dagger [\bm{h}(\bm{k}) \cdot
\bm{\sigma}] U(\bm{k}) = \sigma^3 h(\bm{k})$ 
with $h(\bm{k}) = | \bm{h}(\bm{k})| $ and a new 
spinor $\phi_{\bm{k}} =  U(\bm{k})^\dagger \psi_{\bm{k}} 
=  ( a_{\bm{k} +}, a_{\bm{k} -} )^T$, 
and then obtain
\begin{equation}
U(\bm{k})^\dagger H U(\bm{k})
= \sum_{\bm{k}} \left[
h_{\bm{k}} (n_{\bm{k} +} -n_{\bm{k} -})
+  \frac{1}{2} U ( n_{\bm{k}} - 1)^2
\right],
\end{equation}
with $n_{\bm{k} \pm} = a^\dagger_{\bm{k} \pm} a_{\bm{k} \pm}$ 
and $n_{\bm{k}} = n_{\bm{k} +}+  n_{\bm{k} -}$.
There are four eigenstates and eigenenergies:
(i)  $ \ket{\t{vac}}$ with $E= \frac{U(\bm{k})}{2}$,
(ii)  $a_{\bm{k} +}^\dagger \ket{\t{vac}}$ with $E= h(\bm{k})$,
(iii) $a_{\bm{k} -}^\dagger \ket{\t{vac}}$ with $E=- h(\bm{k})$,
and (iv) $a_{\bm{k} +}^\dagger a_{\bm{k} -}^\dagger \ket{\t{vac}}$ 
with $E=\frac{U(\bm{k})}{2}$.

\begin{figure}[tb]
\begin{center}
\includegraphics[width=1.0\linewidth]{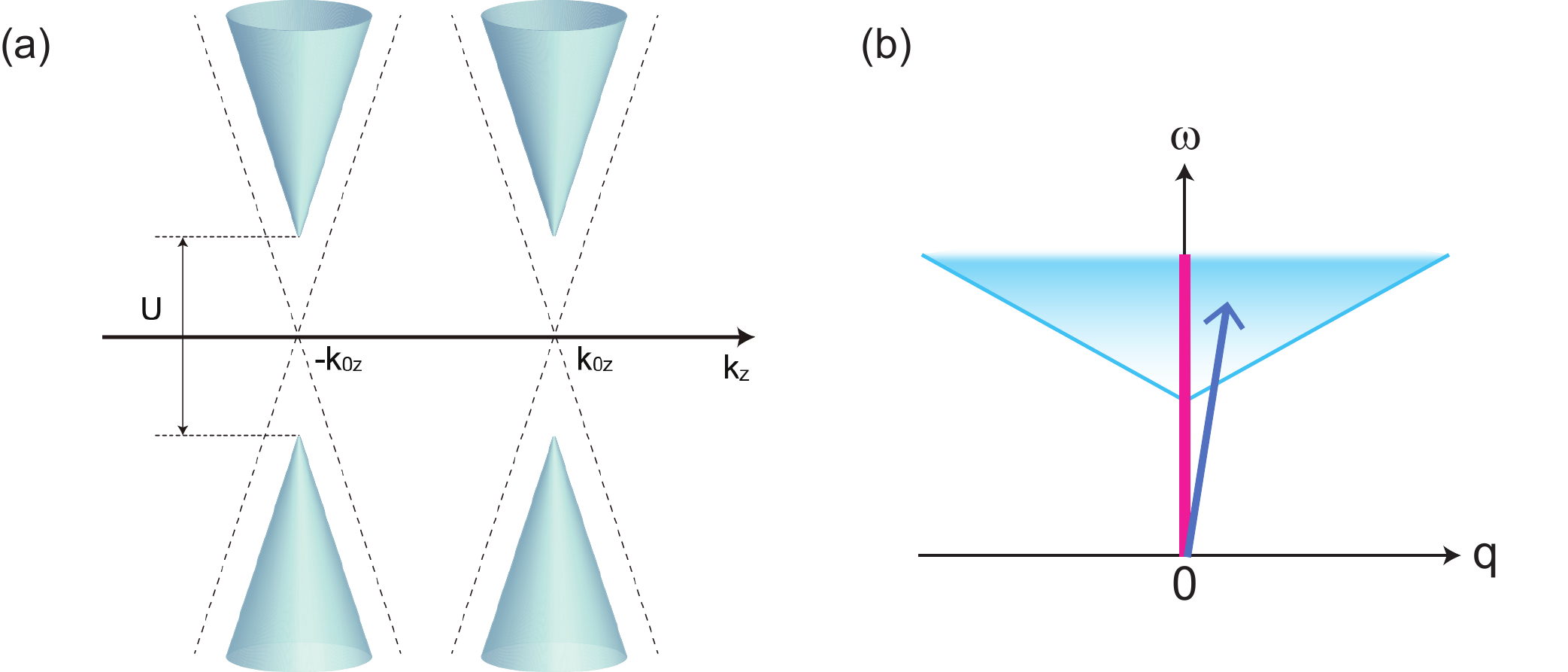}
\end{center}
\caption{
Energy spectrum of WMI. (a) Poles of the Green's function. Dashed line represents 
the band structure of a noninteracting Weyl semimetal. Energy bands are shifted by 
$\pm U/2$ and show a Mott gap of $U$.
(b) Excitation spectrum indicating the region of non-zero conductivity $\sigma(q,\omega)$. 
Excitation gap of $U+v_F |q|$ is finite for nonzero momentum transfer $q$. The spectrum is 
singular at $q=0$ where gapless excitations are allowed due to $k$-local excitations without 
the energy cost of $U$.
}
\label{Fig: band}
\end{figure}

Using these solutions, one can easily obtain the thermal Green's function 
in the zero temperature limit as
\begin{align}
\hat{G}^{-1}(\bm{k}, i \omega) = 
 i \omega \hat{1} - \bm{h}_{\rm eff}(\bm{k}) \cdot
\bm{\sigma} 
\label{eq:GF}
\end{align}
where  $\bm{h}_{\rm eff}(\bm{k}) = 
\bm{n}(\bm{k})[ h(\bm{k}) + U/2 ]  $
with $\bm{n} (\bm{k}) =\bm{h}(\bm{k})/h(\bm{k})$.
(For details, see Supplementary Information SI.)
As can be seen from Eq.~(\ref{eq:GF}),
the energy dispersions of the poles are $\varepsilon_{\pm} (\bm{k}) = \pm
h(\bm{k}) + U/2 $, where the Mott gap of $U$ exists
even at the Weyl point with $h(\bm{k})=0$ as shown in Fig. \ref{Fig: band}(a),
which can be measured in the angle-resolved photoemission spectroscopy (ARPES). 
It should be noticed that Eq.~(\ref{eq:GF}) is derived 
from the exact Green's function and is not a result of some mean-field approximation.
Thus, the WFs disappear due to the electron correlation without the 
pair annihilation. 

\textit{Topological properties ---}
The topological index for the interacting electronic systems can be
defined in terms of Green's function~\cite{Gurarie}.
For the (2+1)D case, it is given by 
\begin{align}
Ch(k_z) &= \frac{\varepsilon_{\alpha \beta \gamma}}{6} \int_{- \infty}^{\infty}
d \omega \int \frac{ d^2 \bm{k} }{(2 \pi)^2} \n
&\quad \times
\t{tr} [ (\hat{G}^{-1} \partial_{k_\alpha} \hat{G})
(\hat{G}^{-1} \partial_{k_\beta} \hat{G})
(\hat{G}^{-1} \partial_{k_\gamma} \hat{G}) ],
\label{eq:N3}
\end{align}
where $\alpha, \beta, \gamma$ run over $0,1,2$,
and $\varepsilon_{\alpha \beta \gamma}$
is the totally antisymmetric tensor.
Plugging Eq.~(\ref{eq:GF}) into Eq.~(\ref{eq:N3}), 
one obtains
\begin{align}
Ch(k_z) = \frac{\varepsilon_{\alpha \beta}}{2} 
\int \frac{ d^2 \bm{k} }{(2 \pi)^2} 
\bm{n}_{\bm{k}} \cdot 
\biggl(
\frac{\partial \bm{n}_{\bm{k}}}{\partial_{k_x}}
\times 
\frac{\partial \bm{n}_{\bm{k}}}{\partial_{k_y}}
\biggr),
\label{eq:Ch}
\end{align}
where the $k$-integral is over $k_x$ and $k_y$ for fixed $k_z$.
Since 
$\bm{n}(\bm{k}) =  \bm{h}_{\rm eff}(\bm{k})/| \bm{h}_{\rm eff}(\bm{k})|
= \bm{h}(\bm{k})/|\bm{h}(\bm{k})|$,
$Ch(k_z)$ does not change in spite of the gap opening at Weyl points.
%This can be understood by considering a small
%surface enclosing the Weyl point $\bm{k}_0$, i.e., $|\bm{h}(\bm{k}_0)|=0$,
%(instead of the plane of fixed $k_z$-plane)
%which continues to act as a monopole even though the WF is gapped. 
The Green's function in Eq.~(\ref{eq:GF}) 
depends on the direction in which
$\bm{k}$ approaches to the Weyl point $\bm{k}_0$ and  
still plays a role of a source (sink) of the Berry curvature. 
At exactly $\bm{k}_0$, $\hat{G}$ has a zero at $\omega=0$ when one averages
over the direction of ${\rm limit}_{\bm{k} \to \bm{k}_0 } \bm{n}(\bm{k})$.
Namely, the role of a pole is replaced by a zero in the topological properties of 
Green's function~\cite{Gurarie}.
Because of the bulk-edge correspondence,
nonzero topological index $Ch(k_z)$
indicates that 
the existence of the surface states on the side surface, i.e., the Fermi arc.
(For details, see Supplementary Information SII.)

%Since the gap opens, the longitudinal conductivity 
%$\sigma_{xx}=5\sigma_{yy} = \sigma_{zz}$ is zero in the DC limit, while
%Hall conductivity $\sigma_{xy}$ remains nonvanishing and is proportional to
%$2 k_{0 z}$, which is given by the $k_z$-integral of $Ch(k_z)$.
Therefore, the present insulating state is topological and we call it 
``Weyl Mott insulator (WMI)'' distinct from 
the usual antiferromagnetic Mott insulator (AFI).

\begin{figure}[tb]
\begin{center}
\includegraphics[width=0.7\linewidth]{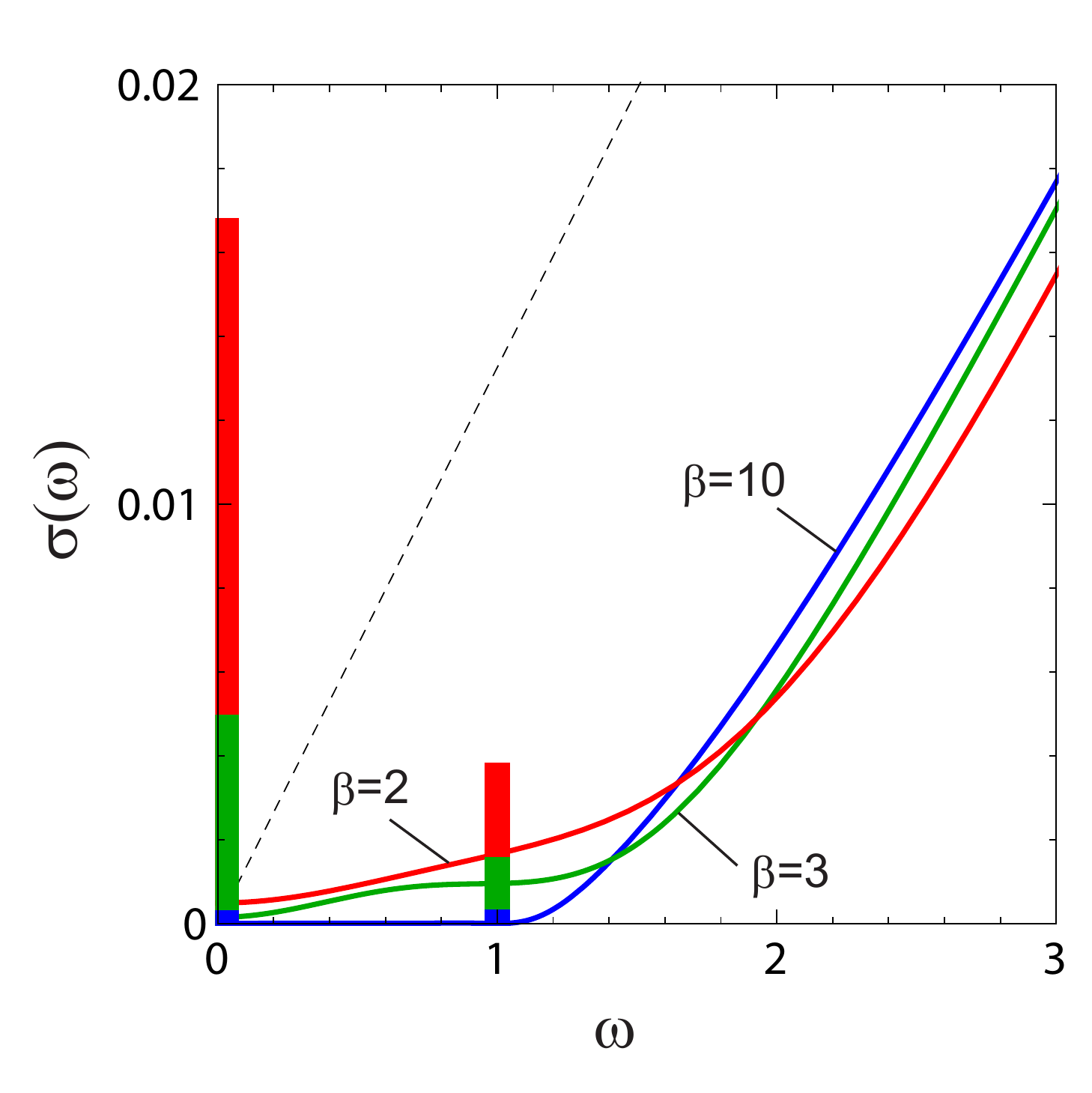}
\end{center}
\caption{
Temperature dependence of the optical conductivity of WMIs.
We plotted results for $\beta=10$ (blue),
$\beta=3$ (green), and 
$\beta=2$ (red).
The delta functions at $\omega=0$ and $U$ from the intraband contributions 
are denoted by bars whose heights represent mutual ratios of the weights.
The dashed line represents $\sigma(\omega)$ for a free WF.
We set $U=1$. 
}
\label{Fig: sxx}
\end{figure}

\textit{Optical conductivity ---}
Now we study the conductivity, which is given by the two-particle correlation function. 
We consider a single WF with the Fermi velocity $v_F$ described by 
$\Vec h(\Vec k)=v_F \Vec k$.
It is crucial to distinguish between the nonzero momentum 
transfer $\bm{q}$ and exactly $\bm{q}=\bm{0}$. 
In the former case, the double occupancy of the electrons will be created, while not in the latter case,
which brings about the singularity or discontinuity at $\bm{q}=\bm{0}$.
This reflects the long-range nature of our Coulomb interaction in 
Eq.~(\ref{eq:Hamiltonian}). 
For $\bm{q} \ne \bm{0}$, the particle-hole continuum starts from 
$\omega = U + v_F |\bm{q}|$ for the transition of an electron from $\bm{k}$ 
to $\bm{k}+\bm{q}$.

For the optical conductivity, the momentum of the incident light $\bm{q}$ is finite although small, 
and hence we take the limit $\bm{q} \to \bm{0}$. 
%(We discuss the case for exactly $\bm{q} = \bm{0}$ later.)
In this limit, the optical conductivity at zero temperature for a single WF 
is obtained as (See Supplementary Information SIII for details)
\begin{align}
\sigma(\Vec q \to \Vec 0,\omega)
=\frac{e^2}{12 h v_F \omega}(\omega-U)^2 \theta(\omega-U),
\label{eq: optical conductivity}
\end{align}
where $\theta(x)=1 (x\ge 0)$ and $\theta(x)=0 (x< 0)$.
%The derivation of $\sigma(\omega)$ both for the zero temperature and the finite temperature is detailed in Supplementary Material.
The optical conductivity shows a Mott gap of $U$ and 
its asymptotic behavior for large $\omega$ is given by
$\sigma(\omega) \simeq (e^2/12hv_F)\omega$ 
which coincides with the well known result for a free WF \cite{Hosur}.
In Fig.~\ref{Fig: sxx}, the optical conductivity is plotted for various temperatures
(See Supplementary Information SIII for the derivation).
As the temperature increases,
peaks at $\omega=0$ and $\omega=U$ appear, as denoted by bars whose heights represent mutual ratios of the weights of the peaks.
The peak at $\omega=0$ is a Drude peak for finite temperatures,
while the peak at $\omega=U$ arises from an intraband contribution in 
which a WF at $\Vec k$ scattered to $\Vec k+\Vec q$ within the same band feels a Coulomb repulsion $U$.
The appearance of the peak at $\omega=U$ in $\sigma(\omega)$ for WMIs contrasts to the absence of such a peak for AFIs,
because the peak at $\omega=U$ originates from the correlation effect.
In addition, the appearance of the in-gap absorption indicates the fragile nature of the Mott gap compared with the single-particle band gap. 

For the case of exactly $\bm{q} = \bm{0}$, 
only the vertical transitions within the same $\bm{k}$-sector contribute to the conductivity as indicated by the red line in Fig.~\ref{Fig: band}(b).
Since no double occupancy is created in this case, no energy cost of $U$ occurs.
Therefore, the conductivity $\sigma(\Vec q = \Vec 0,\omega)$
is given by Eq.~(\ref{eq: optical conductivity}) with $U=0$.

\begin{figure}[tb]
\begin{center}
\includegraphics[width=1.0\linewidth]{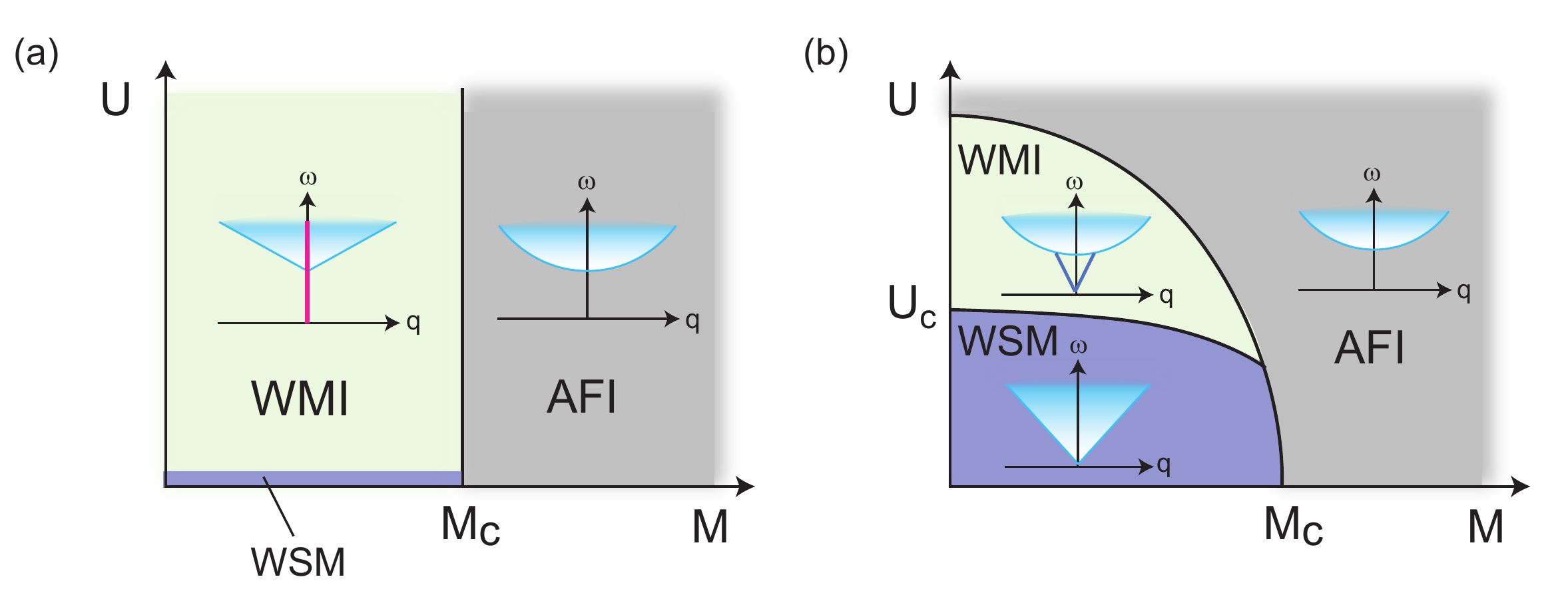}
\end{center}
\caption{
Phase diagrams of the Weyl semimetal. 
(a) The phase diagram of the Hamiltonian in Eq.~(\ref{eq:Hamiltonian}) for WMIs.
(b) The conjectured phase diagram for Weyl semimetals with more realistic interactions.
Insets show the excitation spectra of the conductivity $\sigma(q,\omega)$. 
The singularity at $q=0$ in the excitation spectrum for the WMI phase
turns into a gapless collective excitation for realistic interactions.
}
\label{Fig: phase diagram}
\end{figure}

\textit{Discussion ---}
Now the relevance of the present results to realistic systems is discussed.
There are clear differences between the WMI and the AFI
due to their topological nature: 
(i) The Hall conductance is finite in WMI while it is zero in AFI.
(ii) Correspondingly, the Fermi arc on the surface remains in WMI while not in AFI.
Also importantly, the phase transition between the 
WMI and AFI is possible once the former exists as explained below.
Figure~\ref{Fig: phase diagram}(a) shows the phase diagram of the present model. 
The horizontal axis is the separation $\Delta k_z$ between the two WFs which is controlled by, e.g.,
the strength of the antiferromagnetic long range order parameter $M$.
When $U=0$, the phase transition occurs from the Weyl semimetal to the AFI
by the pair annihilation of WFs at $M=M_c$. 
Once the interaction $U$ is switched on, we always opens the gap and the
system becomes the WMI as long as $\Delta k_z$ is finite. 
Along the phase transition line $\Delta k_z=0$ at $M=M_c$, $U>0$, 
the pair annihilation 
of the two zeros of the Green's function occurs, which is distinct from that 
at $U=0$ where the two poles collide and pair-annihilate.
Here one must consider the peculiarity of the present model.
The effect of the long range Coulomb interactions 
is marginally irrelevant as in the case of quantum electrodynamics (QED)~\cite{Isobe}.
As for the short-range Coulomb interaction $U$, it is
irrelevant. This means that there must be a finite range of Coulomb interaction 
within which the WFs remains gapless and stable. 
On the other hand, the strong $U$ limit in the lattice
model corresponds to the localized electron and hence a trivial AFI. 
Therefore, the conjectured phase diagram of a more realistic model 
is given in Fig.~\ref{Fig: phase diagram}(b), where the successive transitions from the Weyl semimetal to 
the WMI, and from the WMI to the AFI occurs as the strength of the interaction increases.
(The separation of two WFs is reduced also as the interaction increases and hence
the trajectory should goes as $U$ and $M$ simultaneously increase.)

Now we discuss the Green's function and the two-particle correlation function for realistic electron-electron interactions given by
\begin{align}
H_C=\sum_{\bm k,\bm{k'},\bm q} V(\bm{q}) c^\dagger_{\bm{k} +\bm{q}, \sigma} c^\dagger_{\bm{k'} -\bm{q}, \sigma'}
c_{\bm{k'} , \sigma'}c_{\bm{k} , \sigma}.
\label{eq: realistic interaction}
\end{align}
The self-energy $\Sigma(\Vec k=\Vec 0,\omega)$ of the Green's function in the second order in $V(\bm{q})$ is given in Supplementary Information SIV.
It is concluded that the gap of the spectral function is stable and remains nonzero.
The two-particle correlation functions such as $\sigma(\Vec q,\omega)$,
on the other hand,
is gapless at $\Vec q=\Vec 0$ for the Hamiltonian in Eq.~(\ref{eq:Hamiltonian}).
%For Eq.~(\ref{eq: realistic interaction}), the discontinuity at $\Vec q=\Vec 0$ should be removed.
For a finite size system of $N$ sites, the number of poles forming this gapless excitation in the two particle correlation function [the red line in Fig.~\ref{Fig: band}(b)] is of $O(N)$ (which is the number of $\Vec k$ where the excitation can be created). 
In general, the number of poles for a collective excitation is of $O(N)$, while that of a continuous excitation arising as a pair of single-particle excitations is of $O(N^2)$.
This infers that the gapless excitations at $\Vec q = \Vec 0$ corresponds to a collective excitation.
For realistic interactions, the discontinuity at $\bm{q}=\bm{0}$ should be removed. 
In this case, we conjecture that the vertical transition 
[red line in the inset of Fig.~\ref{Fig: phase diagram}(a)] turns into a collective mode with a linear dispersion as shown in the blue line in the inset in the WMI phase of Fig.~\ref{Fig: phase diagram}(b). 

These considerations offer a different scenario to interpret the phase diagram of 
pyrochlore iridates $R_2$Ir$_2$O$_7$~\cite{Wan11,WWK12}.
As the radius of the rare-earth ion $R$
is reduced, the correlation strength increases. 
A recent optical measurement in 
Nd$_2$Ir$_2$O$_7$ has revealed the opening of the Mott gap of the order of 
0.05eV~\cite{Ueda1}. A transport experiment also discovered the metallic 
domain wall states even in the Mott insulating state, i.e., the bulk is insulating 
while the domain wall is metallic~\cite{Ueda2}. As the correlation is further 
reduced, these surface metallic states also disappear. One scenario is proposed by 
Yamaji et al.~\cite{Yamaji} based on the mean field theory. As an alternative 
scenario, one can consider the two types of Mott insulators, 
i.e., the WMI and the AFI,
and the disappearance of the metallic domain wall states signals the phase 
transition between the two phases.
Namely, since two domains of the antiferromagnetic order
correspond to opposite signs of $\sigma_{xy}$ and hence the 
two-dimensional chiral surface modes are expected to appear
at the domain boundary in the WMI phase.
However, we note that this requires a symmetry lowering to violate the 
cancellation of the Chern vectors pointing toward four momentum directions
equivalent to (1,1,1) which makes $\sigma_{xy}$ zero in the cubic symmetric case \cite{Yang}.
The smoking-gun experiment for the WMI should be the ARPES to detect the 
Fermi arc even in the Mott insulating phase as mentioned above.

\textit{Acknowledgment ---}
We thank Y. Tokura, B.-J. Yang, L. Balents, L. Fu, and W. Witczak-Krempa for fruitful discussions.
This work was supported by 
Grant-in-Aid for Scientific Research 
(No.~24224009, and No.~26103006)
from the Ministry 
of Education, Culture, Sports, Science and
Technology (MEXT) of Japan
and from Japan Society for the Promotion of Science.

\bibliographystyle{naturemag.bst}
\bibliography{WMI}

%Start Supplemental Material

%%%%%%%%%% Merge with supplemental materials %%%%%%%%%%
%\pagebreak
\clearpage
\widetext
\begin{center}
\textbf{\large Supplementary Information for 
``Weyl Mott insulator''}
\end{center}

%%%%%%%%%% Prefix a "S" to all equations, figures, tables and reset the counter %%%%%%%%%%
\setcounter{equation}{0}
\setcounter{figure}{0}
\setcounter{table}{0}
\setcounter{section}{0}
\makeatletter
\renewcommand{\thesection}{S\Roman{section}}
\renewcommand{\theequation}{S\arabic{equation}}
\renewcommand{\thefigure}{S\arabic{figure}}
\renewcommand{\bibnumfmt}[1]{[S#1]}
\renewcommand{\citenumfont}[1]{S#1}
%%%%%%%%%% Prefix a "S" to all equations, figures, tables and reset the counter %%%%%%%%%%

\section{Green's function}
We derive the Green's function for the Hamiltonian given by 
\begin{align}
H=\psi^\dagger_{\Vec{k}} \Vec{h}(\Vec k)\cdot \Vec{\sigma} \psi_{\Vec{k}}
+
\frac 1 2 U (n_{\Vec{k}}-1)(n_{\Vec{k}}-1),
\label{eq: H}
\end{align}
where
\begin{align}
\psi_{\Vec k} &=
\begin{pmatrix}
c_{\Vec k \uparrow} \\
c_{\Vec k \downarrow}
\end{pmatrix},
\end{align}
$\sigma$ are Pauli matrices acting on spin degrees of freedom,
$n_{\Vec{k}}$ is the density operator 
$n_{\Vec{k}}=\psi^\dagger_{\Vec{k}}\psi^{\,}_{\Vec{k}}$,
and $U$ is the magnitude of the repulsive interaction.
The repulsive interaction is infinite-ranged in the real space,
which can be represented in a local way in the momentum space.
Thanks to the locality in the momentum space, 
the Green's function can be exactly computed for this Hamiltonian as follows.

First we perform a unitary transformation
\begin{align}
\begin{pmatrix}
c_{\Vec k \uparrow} \\
c_{\Vec k \downarrow}
\end{pmatrix}
&=
U(\Vec k)
\begin{pmatrix}
b_{\Vec k +} \\
b_{\Vec k -}
\end{pmatrix}
\end{align}
that diagonalizes the single-particle part of the Hamiltonian as
\begin{align}
U^\dagger(\Vec k) [\Vec{h}(\Vec k)\cdot \Vec \sigma]U(\Vec k)
&=
h(\Vec k)\sigma_z.
\end{align}
Then the Green's function is transformed as
\begin{align}
G_{\alpha \beta}(\Vec k,\tau) &\equiv 
-\braket{ T_\tau c_{\Vec k\alpha}(\tau) c_{\Vec k\beta}^\dagger}
=-U_{\alpha a}(\Vec k)U_{\beta a'}(\Vec k)
\braket{ T_\tau b_{\Vec ka}(\tau) b_{\Vec ka'}^\dagger},
\end{align}
where $T_\tau$ denotes the time ordering and 
$c_{\Vec k\alpha}(\tau)=e^{\tau H} c_{\Vec k\alpha}e^{-\tau H}$.

In this basis, the Hamiltonian is diagonalized as
\begin{align}
H\ket{0}&=\frac{U}{2}\ket{0}, \\
Hb_{\Vec k-}^\dagger\ket{0}&=-h(\Vec k)b_{\Vec k-}^\dagger\ket{0}, \\
Hb_{\Vec k+}^\dagger\ket{0}&=h(\Vec k)b_{\Vec k+}^\dagger\ket{0}, \\
Hb_{\Vec k-}^\dagger b_{\Vec k+}^\dagger\ket{0}&=\frac{U}{2}b_{\Vec k-}^\dagger b_{\Vec k+}^\dagger\ket{0}, 
\end{align}
where $\ket 0$ is the vacuum state and $h(\Vec k)=|\Vec h(\Vec k)|$. 
Thus the expectation value is written as
\begin{align}
\braket{ b_{\Vec ka}(\tau) b_{\Vec ka'}^\dagger}
&=
\frac{1}{Z}[
\bra{0} b_{\Vec ka}(\tau) b_{\Vec ka'}^\dagger \ket{0} e^{-\beta \frac U 2} \n
&\qquad + \bra{0} b_{\Vec k-} b_{\Vec ka}(\tau) b_{ka'}^\dagger b_{\Vec k-}^\dagger \ket{0} e^{\beta h(\Vec k)} \n
&\qquad + \bra{0} b_{\Vec k+} b_{\Vec ka}(\tau) b_{\Vec ka'}^\dagger b_{\Vec k+}^\dagger \ket{0} e^{-\beta h(\Vec k)} \n
&\qquad + \bra{0} b_{\Vec k+}b_{\Vec k-} b_{ka}(\tau) b_{\Vec ka'}^\dagger b_{\Vec k-}^\dagger b_{\Vec k+}^\dagger \ket{0} e^{-\beta \frac U 2}  ], \\
Z&=e^{\beta h(\Vec k)}+e^{-\beta h(\Vec k)}+2 e^{-\beta \frac U 2}.
\end{align}
In the right hand side of the equation for $\braket{ b_{\Vec ka}(\tau) b_{\Vec ka'}^\dagger}$, the forth term vanishes 
and other terms are nonzero when $a=a'$.
Thus we obtain
\begin{align}
\braket{b_{\Vec k+}(\tau) b_{\Vec k+}^\dagger}&=
\frac{1}{Z}
(\bra{0} b_{\Vec k+}(\tau) b_{\Vec k+}^\dagger \ket{0} e^{-\beta \frac U 2} 
+ \bra{0} b_{\Vec k-} b_{\Vec k+}(\tau) b_{\Vec k+}^\dagger b_{\Vec k-}^\dagger \ket{0} e^{\beta h(\Vec k)}) \n
&=
\frac{1}{Z}(e^{\tau(-h+\frac{U}{2})-\beta\frac{U}{2}}
+e^{\tau(-h-\frac{U}{2})+\beta h}),
\\
\braket{b_{\Vec k-}(\tau) b_{\Vec k-}^\dagger}&=
\frac{1}{Z}
(\bra{0} b_{\Vec k-}(\tau) b_{\Vec k-}^\dagger \ket{0} e^{-\beta \frac U 2} 
+ \bra{0} b_{\Vec k+} b_{\Vec k-}(\tau) b_{\Vec k-}^\dagger b_{\Vec k+}^\dagger \ket{0} e^{-\beta h(\Vec k)}) \n
&=
\frac{1}{Z}(e^{\tau(h+\frac{U}{2})-\beta\frac{U}{2}}
+e^{\tau(h-\frac{U}{2})-\beta h}).
\end{align}

Therefore, the imaginary-time Green's function is given by
\begin{align}
G_{++}(\Vec k,i\omega_n) &= -\int_0^\beta d\tau e^{i\omega_n \tau} \braket{b_{\Vec k+}(\tau) b_{\Vec k+}^\dagger} \n
&=\frac{1}{Z}\left(
\frac{e^{-\beta h}+e^{-\beta \frac U 2}}{i\omega_n-h+\frac U 2}
+\frac{e^{\beta h}+e^{-\beta \frac U 2}}{i\omega_n-h-\frac U 2}
\right),
\\
G_{--}(\Vec k,i\omega_n) &= -\int_0^\beta d\tau e^{i\omega_n \tau} \braket{b_{\Vec k-}(\tau) b_{\Vec k-}^\dagger} \n
&=\frac{1}{Z}\left(
\frac{e^{-\beta h}+e^{-\beta \frac U 2}}{i\omega_n+h-\frac U 2}
+\frac{e^{\beta h}+e^{-\beta \frac U 2}}{i\omega_n+h+\frac U 2}
\right).
\end{align}

\subsection{Green's function for $T=0$}
In the zero temperature ($\beta \to \infinity$),
the above Green' function reduces to
\begin{align}
G_{++}(\Vec k,i\omega_n) &=\frac{1}{i\omega_n-h-\frac U 2},\\
G_{--}(\Vec k,i\omega_n) &=\frac{1}{i\omega_n+h+\frac U 2}.
\end{align}

In the original basis, the Green's function is given by
\begin{align}
G(\Vec k,i\omega_n) &=\frac{1}{i\omega_n-\left(h+\frac U 2\right)\Vec n \cdot \Vec \sigma}, &
\Vec n&= \frac{\Vec h}{h}.
\end{align}

\section{Fermi arc}
In this section, we study the Fermi arc in the WMIs.
Nonvanishing topological indices for the WMIs
indicate that the Fermi arc remains in the WMIs,
which we verify in the following.
%the existence of the surface states on the side surface, i.e., the Fermi arc.
Since our model [Eq.~(\ref{eq: H})] is diagonalized at each $\bm{k}$-point
and hence we can consider an effective two-dimensional model 
for each $k_z$-sector 
when the system is periodic in the $z$-direction.
%for an infinite cylinder along $z$-direction.
Because of the bulk-boundary correspondence, we expect 
that ``edge channels'' for each $k_z$ form a Fermi arc.
More explicitly, 
one can obtain the surface bound state from 
the effective Hamiltonian 
\begin{align}
H_\t{eff}&=
\left[ h(\bm{k}) + \frac U 2 \right]
\bm{n}(\bm{k})\cdot \bm{\sigma},  
&
\bm{h}(\bm{k})&=v_F(k_x,k_y,k_z),
\label{eq:effective}
\end{align}
by replacing the momenta $k_x, k_y$ 
with the derivatives $-i \partial_x, -i \partial_y$. 
Away from the plane $k_z = \pm k_{0z}$,
the surface state is almost unchanged from the noninteracting case.
The nontrivial issue is how the surface state behaves as $k_z$ approaches $\pm k_{0z}$.
Specifically, the problem is whether the
penetration depth of the surface states diverges or not with
$k_z \to \pm k_{0z}$.
Intuitively, the finite gap $U$ indicates that the length scale $\xi$
remains finite, i.e., $\xi \cong \hbar v_F/ U$.
%where $v_F$ is the velocity in Eq.~(\ref{eq:WF}).
However, it turns out not when one studies the effective Hamiltonian in Eq.~(\ref{eq:effective}) and the asymptotic behavior of 
the surface bound state as $|x| \to \infty$ (here we assume $k_y=0$)
by tentatively taking the limit of $|k_x| \ll |k_z|$.
In this limit, 
$H_{\rm eff} \cong ( v_F + \frac{U}{2} |k_z|^{-1} ) [ -i \partial_x \sigma^1 + k_z \sigma^3]$,
which indicates that the penetration depth diverges with 
$\xi = |k_z|^{-1}$. 
In any case, the length scale is determined by $|k_z|^{-1}$ even when we take into account of the higher orders in $\partial_x$.  
Therefore, the surface bound states penetrate into the bulk
as $k_z$ approaches to $\pm k_{0z}$.

\section{Optical conductivity}
We study the optical conductivity $\sigma(\omega)$ 
for a single WF described $\Vec h(\Vec k)= v_F \Vec k$.
In the following, we set the Fermi velocity $v_F=1$,
which can be always restored by the dimension analysis.

\subsection{Matrix elements}
Here we calculate matrix elements that we will need in evaluation of conductivities, i.e., $\bra \pm \sigma_i \ket \pm$.
We first parameterize the direction of the momentum as
\begin{align}
\Vec n=(\sin \theta \cos \phi, \sin \theta \cos \phi, \cos \theta).
\end{align}
Then the wave functions that diagonalize the Hamiltonian are written as
\begin{align}
\ket + &=
\begin{pmatrix}
\cos \frac{\theta}{2} \\
e^{i\phi} \sin \frac{\theta}{2}
\end{pmatrix},
&
\ket - &=
\begin{pmatrix}
-\sin \frac{\theta}{2} \\
e^{i\phi} \cos \frac{\theta}{2}
\end{pmatrix}.
\end{align}
The matrix elements are given by
\begin{align}
\bra + \sigma_x \ket + &=
\sin \theta \cos \phi , \\
\bra - \sigma_x \ket - &=
-\sin \theta \cos \phi , \\
\bra + \sigma_x \ket - &=
\cos \theta \cos \phi+ i \sin \phi, \\
\bra + \sigma_y \ket + &=
\sin \theta \sin \phi , \\
\bra + \sigma_y \ket - &=
-i \cos \phi + \cos \theta \sin \phi . 
\end{align}
In the evaluation of the optical conductivity, we need
\begin{align}
\int \sin \theta d\theta d\phi 
\bra \pm \sigma_x \ket \pm \bra \pm \sigma_x \ket \pm
&=\frac {4\pi}{3}, \\
\int \sin \theta d\theta d\phi 
\bra + \sigma_x \ket - \bra - \sigma_x \ket +
&=\frac {8\pi}{3}.
\end{align}
In the evaluation of the Hall conductivity as a function of $k_z$, we need
\begin{align}
\int d\phi 
\bra + \sigma_x \ket + \bra + \sigma_y \ket +
&=0, \\
\int d\phi 
\bra + \sigma_x \ket - \bra - \sigma_y \ket +
&=2\pi i \cos \theta=2\pi i \frac{k_z}{k}. 
\label{eq: matrix elements for sxy}
\end{align}

\subsection{Zero temperature}
We first focus on the conductivity $\sigma(\omega)$ for the zero temperature.
The Green's function is given by
\begin{align}
G(i\omega_m)&=
\frac{1}{(i\omega_m)^2-(k+\frac{U}{2})^2} 
\left[i\omega_m + \left(k+\frac{U}{2}\right) \Vec n \cdot \Vec \sigma \right],
\end{align}
with $\Vec n=\Vec k/|\Vec k|$.
The optical conductivity is given by
\begin{align}
\sigma(\omega)&=\t{Re}\left[\frac{Q(\omega+i\epsilon)}{-i\omega}\right], \\
Q(i\Omega)&=
\lim_{\Vec q \to \Vec 0}
\int \frac{d^3\Vec k}{(2\pi)^3} \sum_{i\omega_m} 
\t{tr}[G(\Vec k, i\omega_m)\sigma_x G(\Vec k+ \Vec q, i\omega_m+i\Omega)\sigma_x].
\end{align}

The integrand of $Q(i\Omega)$ reads
\begin{align}
&\sum_{i\omega_m} 
\t{tr}[G(\Vec k, i\omega_m)\sigma_x G(\Vec k, i\omega_m+i\Omega)\sigma_x]
\n
&=
\sum_{i\omega_m}
\frac{1}{(i\omega_m+i\Omega)^2-(k+\frac{U}{2})^2}
\frac{1}{(i\omega_m)^2-(k+\frac{U}{2})^2} \n
&\qquad \t{tr}\left[
\left(i\omega_m+i\Omega + \left(k+\frac{U}{2}\right) \Vec n \cdot \Vec \sigma \right)
\sigma_x
\left(i\omega_m + \left(k+\frac{U}{2}\right) \Vec n \cdot \Vec \sigma \right)
\sigma_x
\right] \n
&=\sum_{i\omega_m}
\frac{2}{(i\omega_m+i\Omega)^2-(k+\frac{U}{2})^2}
\frac{1}{(i\omega_m)^2-(k+\frac{U}{2})^2}
\left[(i\omega_m+i\Omega)i\omega_m + \left(1+\frac{U}{2k}\right)^2 (k_x^2-k_y^2-k_z^2)
\right] \n
&=\sum_{i\omega_m}
\frac{2}{(i\omega_m+i\Omega)^2-(k+\frac{U}{2})^2}
\frac{1}{(i\omega_m)^2-(k+\frac{U}{2})^2}
\left[(i\omega_m+i\Omega)i\omega_m -\frac{1}{3} \left(k+\frac{U}{2}\right)^2 
\right].
\end{align}
By using the formula
\begin{align}
\sum_{i\omega_m}\frac{\left[(i\omega_m+i\Omega)i\omega_m -abc \right]}
{[(i\omega_m+i\Omega)^2-a^2] [(i\omega_m)^2-b^2]}
&=
\frac{1}{2}
\frac{(a+b)(1-c)}{[i\Omega-(a+b)][i\Omega+(a+b)]}
\label{eq: w sum formula}
\end{align}
for $a>0,b>0$, 
we perform the summation over $i\omega_m$ for the above equation 
and obtain
\begin{align}
\sum_{i\omega_m} 
\t{tr}[G(\Vec k, i\omega_m)\sigma_x G(\Vec k, i\omega_m+i\Omega)\sigma_x]
%\n
&=
\frac{8}{3}
\frac{\left(k+\frac{U}{2}\right)}{[i\Omega-(2k+U)][i\Omega+(2k+U)]}.
\end{align}
After the analytic continuation $i\Omega \to \omega+i\epsilon$,
only the pole at $k=\omega-\frac{U}{2}$ contributes to the imaginary part of the $k$-integral.
Thus, we obtain
\begin{align}
\t{Im}[Q(\omega)]&=
\frac{4}{3\pi^2} \int_0^\infinity k^2 dk 
\frac{\left(k+\frac{U}{2}\right)}{\omega+(2k+U)}
\t{Im}\left[\frac{1}{\omega+i\epsilon-(2k+U)}\right] 
\n
&=-\frac{1}{24\pi} (\omega-U)^2 \theta(\omega-U),
\end{align}
where we used the formula $\t{Im}\left[\frac{1}{k-a-i \epsilon}\right]=\pi \delta(a)$.
Hence, the optical conductivity for the zero temperature is given by
\begin{align}
\sigma(\omega)=-\frac{\t{Im}[Q(\omega)]}{\omega}
=\frac{1}{24 \pi \omega}
\left(\omega-U\right)^2
\theta\left(\omega-U\right).
\end{align}
By restoring the unit of $e^2/\hbar$ and the Fermi velocity $v_F$,
we end up with
\begin{align}
\sigma(\omega)=\frac{e^2}{12 h v_F \omega}
\left(\omega-U\right)^2
\theta\left(\omega-U\right).
\end{align}

\subsubsection{Poles of $\sigma(\Vec q ,\omega)$}
Let us study the locus of the poles of the two-particle correlation function 
that contribute to
the conductivity $\sigma(\Vec q ,\omega)$ for nonzero $\Vec q$.
From Eq.~(\ref{eq: w sum formula}) and setting 
$a=|\Vec k|+\frac U 2$ and $b=|\Vec k+ \Vec q|+\frac U 2$,
the poles of
$\sum_{i\omega_m} 
\t{tr}[G(\Vec k, i\omega_m)\sigma_x G(\Vec k+ \Vec q, i\omega_m+i\Omega)\sigma_x]
$ 
can be read off as $\omega=a+b=|\Vec k|+|\Vec k+ \Vec q|+ U$.
By using the formula $|\Vec k|+|\Vec k+ \Vec q| \ge |\Vec q|$ and restoring the Fermi velocity $v_F$,
the lower bound of the poles is given by
\begin{align}
\omega=U+v_F|\Vec q|.
\end{align}

\subsection{Finite temperature}

In this section, we calculate the optical conductivity $\sigma(\omega)$ in the finite temperature. In doing so, we consider contributions from interband and intraband transitions separately as
\begin{align}
\sigma(\omega)=\sigma^\t{inter}(\omega)+\sigma^\t{intra}(\omega).
\end{align}

\subsubsection{Interband transition}

The interband contribution to $Q(i\Omega)$ is given by
\begin{align}
Q_\t{inter}(i\Omega)=\frac{1}{(2\pi)^3}\int k^2dk A_\t{inter}(i\Omega),
\end{align}
where
\begin{align}
&A_\t{inter}(i\Omega) \n
&\equiv
\int \sin\theta d\theta d\phi 
\sum_{i\omega_m}
[
G_{++}(\Vec k, i\omega_m) \bra{+} \sigma_x \ket{-}
G_{--}(\Vec k, i\omega_m+i\Omega) \bra{-} \sigma_x \ket{+}
\n
&\hspace{8em}
+
G_{++}(\Vec k, i\omega_m+i\Omega) \bra{+} \sigma_x \ket{-}
G_{--}(\Vec k, i\omega_m) \bra{-} \sigma_x \ket{+}
]
\n
&=
\frac{8\pi}{3} Z^{-2}\sum_{s=\pm 1}
\Big[
\left(
\frac{e^{\beta k}+e^{-\beta\frac{U}{2}}}{2k+U+is\Omega}
+
\frac{e^{-\beta k}+e^{-\beta\frac{U}{2}}}{2k+is\Omega}
\right)
e^{-\beta\frac{U}{2}}
%\n
%&\hspace{3em}
+
\left(
\frac{e^{\beta k}+e^{-\beta\frac{U}{2}}}{2k+is\Omega}
+
\frac{e^{-\beta k}+e^{-\beta\frac{U}{2}}}{2k-U+is\Omega}
\right)
e^{-\beta k} 
\n
&\hspace{3em}
+
\left(
\frac{e^{\beta k}+e^{-\beta\frac{U}{2}}}{-(2k+U)+is\Omega}
+
\frac{e^{-\beta k}+e^{-\beta\frac{U}{2}}}{-2k+is\Omega}
\right)
e^{\beta k}
%\n
%&\hspace{3em}
+
\left(
\frac{e^{\beta k}+e^{-\beta\frac{U}{2}}}{-2k+is\Omega}
+
\frac{e^{-\beta k}+e^{-\beta\frac{U}{2}}}{-(2k-U)+is\Omega}
\right)
e^{-\beta \frac{U}{2}}
\Big].
\end{align}
%where we have performed an integration over directions $\int \sin\theta d\theta d\phi$ for current matrix elements.
This is reduced to
\begin{align}
Q_\t{inter}(\omega)&=
\frac{1}{3\pi^2}Z^{-2}\sum_{s=\pm 1}
\int k^2dk
\left(
\frac{e^{\beta k}+e^{-\beta\frac{U}{2}}}{2k+U+is\Omega}
+
\frac{e^{-\beta k}+e^{-\beta\frac{U}{2}}}{2k+is\Omega}
\right)
(-e^{\beta k}+e^{-\beta\frac{U}{2}})
\n
&\hspace{6em}
+
\left(
\frac{e^{\beta k}+e^{-\beta\frac{U}{2}}}{2k+is\Omega}
+
\frac{e^{-\beta k}+e^{-\beta\frac{U}{2}}}{2k-U+is\Omega}
\right)
(e^{-\beta k}-e^{-\beta\frac{U}{2}}).
\end{align}
After the analytic continuation $i\Omega \to \omega+i\epsilon$,
poles that contribute to the imaginary part of 
the $k$-integral
are 
\begin{align}
k=
\frac{\omega-U}{2}, \frac{\omega}{2}, 
\frac{\omega+U}{2}, \frac{U-\omega}{2}.
\end{align}
Thus the interband contribution to the optical conductivity at the finite temperature is given by
\begin{align}
\sigma^\t{inter}(\omega)&=\t{Im}\left[\frac{Q_\t{inter}(\omega)}{-i\omega} \right] \n
&=
-\frac{1}{6\pi\omega}\Big[
(e^{\beta \frac{\omega-U}{2}}+e^{-\beta\frac{U}{2}})(-e^{\beta \frac{\omega-U}{2}}+e^{-\beta\frac{U}{2}})\left(\frac{\omega-U}{2}\right)^2 Z\left(\frac{\omega-U}{2}\right)^{-2} \theta(\omega-U)
\n
&\quad +
2(e^{-\beta \frac{\omega+U}{2}}-e^{\beta \frac{\omega-U}{2}})\left(\frac{\omega}{2}\right)^2 Z\left(\frac{\omega}{2}\right)^{-2}
\n
&\quad +
(e^{-\beta \frac{\omega+U}{2}}+e^{-\beta\frac{U}{2}})(e^{-\beta \frac{\omega+U}{2}}-e^{-\beta\frac{U}{2}})\left(\frac{\omega+U}{2}\right)^2 Z\left(\frac{\omega+U}{2}\right)^{-2} 
\n
&\quad -
(e^{-\beta \frac{-\omega+U}{2}}+e^{-\beta\frac{U}{2}})(e^{-\beta \frac{-\omega+U}{2}}-e^{-\beta\frac{U}{2}})\left(\frac{-\omega+U}{2}\right)^2 Z\left(\frac{-\omega+U}{2}\right)^{-2} \theta(-\omega+U)
\Big]
. 
\end{align}
The minus sign for the term in the last line arises because the pole 
$k=\frac{U-\omega-i\epsilon}{2}$ locates in the lower half plane 
while other poles locate in the upper half plane.

\subsubsection{Intraband transition}

The intraband contribution to $Q(i\Omega)$ is given by
\begin{align}
Q_\t{intra}(i\Omega)=\frac{1}{(2\pi)^3}\int k^2dk 
[A_\t{intra}(i\Omega)+B_\t{intra}(i\Omega)],
\end{align}
where
\begin{align}
A_\t{intra}(i\Omega)&\equiv
\int \sin \theta d\theta d\phi
\sum_{i\omega_n}
G_{++}(k,i\omega_n) \bra{+} \sigma_x \ket{+}
G_{++}(k+q,i\omega_n+i\Omega_m) \bra{+} \sigma_x \ket{+} \n
&=\frac{4\pi}{3}\frac{1}{Z^2}\Big[
(e^{-\beta h}+e^{-\beta \frac U 2})^2\frac{n_F(h_k-\frac U 2)-n_F(h_{k+q}-\frac U 2)}{i\Omega+h_k-h_{k+q}} \n
&\quad +
(e^{\beta h}+e^{-\beta \frac U 2})(e^{-\beta h}+e^{-\beta \frac U 2})\frac{n_F(h_k-\frac U 2)-n_F(h_{k+q}+\frac U 2)}{i\Omega+h_k-h_{k+q}-U} \n
&\quad +
(e^{\beta h}+e^{-\beta \frac U 2})^2\frac{n_F(h_k+\frac U 2)-n_F(h_{k+q}+\frac U 2)}{i\Omega+h_k-h_{k+q}} \n
&\quad +
(e^{\beta h}+e^{-\beta \frac U 2})(e^{-\beta h}+e^{-\beta \frac U 2})\frac{n_F(h_k+\frac U 2)-n_F(h_{k+q}-\frac U 2)}{i\Omega+h_k-h_{k+q}+U}
\Big],
\end{align}
and
\begin{align}
B_\t{intra}(i\Omega)&\equiv
\int \sin \theta d\theta d\phi
\sum_{i\omega_n}G_{--}(k,i\omega_n) \bra{-} \sigma_x \ket{-}
G_{--}(k+q,i\omega_n+i\Omega_m) \bra{-} \sigma_x \ket{-} \n
&=\frac{4\pi}{3}\frac{1}{Z^2}\Big[
(e^{\beta h}+e^{-\beta \frac U 2})^2\frac{n_F(-h_k-\frac U 2)-n_F(-h_{k+q}-\frac U 2)}{i\Omega-h_k+h_{k+q}} \n
&\quad +
(e^{\beta h}+e^{-\beta \frac U 2})(e^{-\beta h}+e^{-\beta \frac U 2})\frac{n_F(-h_k-\frac U 2)-n_F(-h_{k+q}+\frac U 2)}{i\Omega-h_k+h_{k+q}-U} \n
&\quad +
(e^{-\beta h}+e^{-\beta \frac U 2})^2\frac{n_F(-h_k+\frac U 2)-n_F(-h_{k+q}+\frac U 2)}{i\Omega-h_k+h_{k+q}} \n
&\quad +
(e^{\beta h}+e^{-\beta \frac U 2})(e^{-\beta h}+e^{-\beta \frac U 2})\frac{n_F(-h_k+\frac U 2)-n_F(-h_{k+q}-\frac U 2)}{i\Omega-h_k+h_{k+q}+U}
\Big].
\label{eq: intraband G++ G--}
\end{align}
After performing an analytic continuation $i\Omega_m \to \omega + i\epsilon$ and taking a limit $q \to 0$,
we obtain the intraband contribution to the optical conductivity
\begin{align}
&\sigma^\t{intra}(\omega) \n
&=
\frac{1}{6\pi^2}\int dk k^2 \frac{1}{Z(k)^2} \left\{
-(e^{-\beta h}+e^{-\beta \frac U 2})^2 n'_F\left(h_k-\frac U 2\right)
-(e^{\beta h}+e^{-\beta \frac U 2})^2 n'_F\left(h_k+\frac U 2\right)
\right\}\delta(\omega)
\n
&\quad 
+\frac{1}{6\pi^2}\int dk k^2 \frac{1}{Z(k)^2} (e^{\beta h}+e^{-\beta \frac U 2})(e^{-\beta h}+e^{-\beta \frac U 2}) \n
& \hspace{3em} \times
\frac{1}{U}\left[n_F\left(h_k-\frac U 2\right)-n_F\left(h_k+\frac U 2\right)+n_F\left(-h_k-\frac U 2\right)-n_F\left(-h_k+\frac U 2\right)\right]
\delta(\omega-U)
.
\label{eq: sxx intra}
\end{align}
We note that we used the equation $n'_F(\epsilon)=n'_F(-\epsilon)$ in the first term, and the forth term in Eq.~(\ref{eq: intraband G++ G--}) 
can be discarded after analytic continuation because of a factor $\delta(\omega + U)$.

\begin{figure}
\begin{center}
\includegraphics[width=0.5\linewidth]{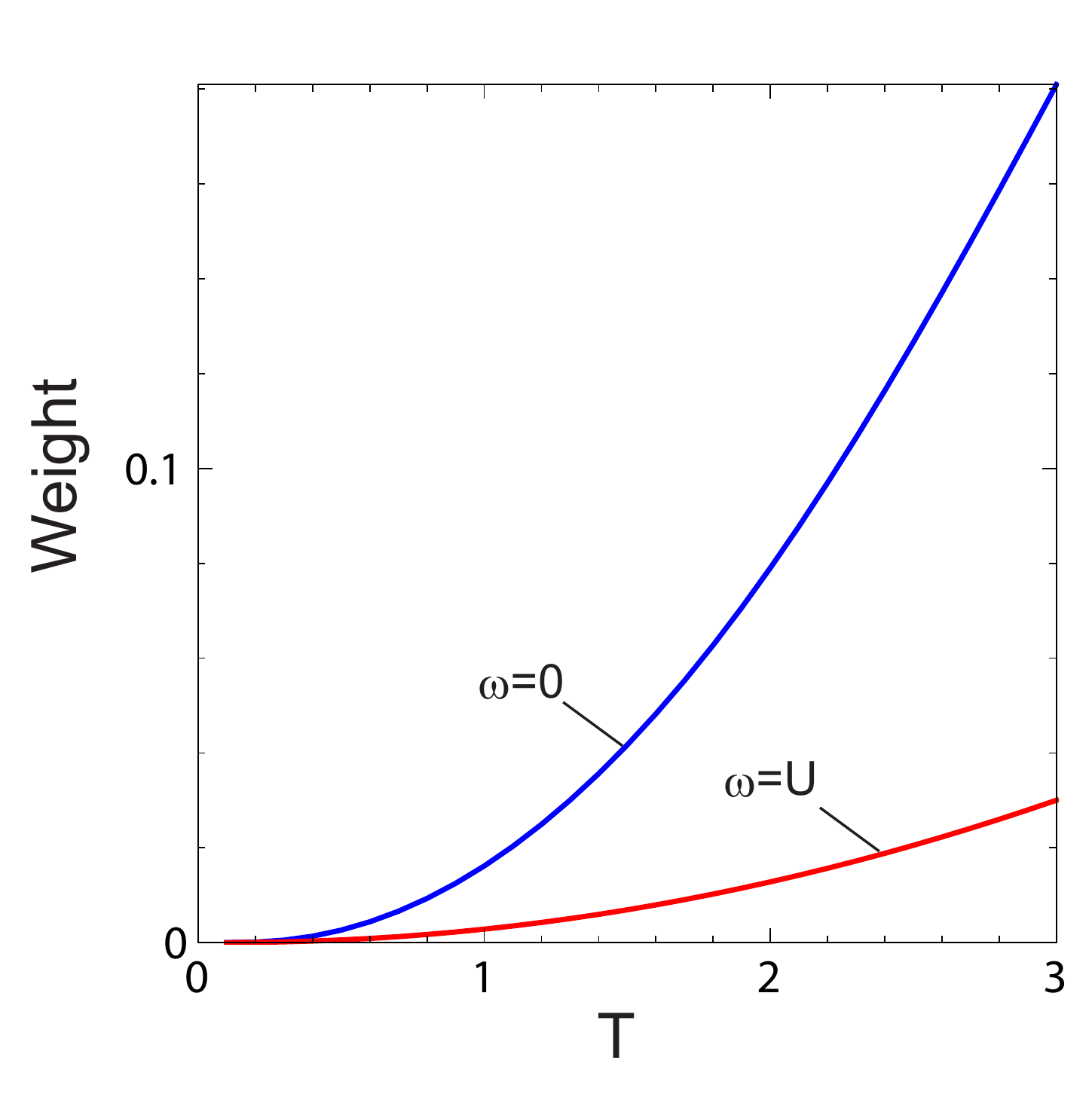}
\caption{\label{fig: T dep of intraband}
Temperature dependence of weights of intraband transitions at $\omega=0$ (blue) and $\omega=U$ (red).}
\end{center}
\end{figure}

%\begin{figure}
%\begin{center}
%\includegraphics[width=0.5\linewidth]{figs/optical-conductivity.pdf}
%\caption{Optical conductivity for $\beta=10$ (blue), $\beta=2$ (green), and $\beta=1$ (red).}
%\end{center}
%\label{fig: sxx inter+intra}
%\end{figure}

We show the temperature dependence of weights of peaks at $\omega=0$ and $U$ in Fig.~\ref{fig: T dep of intraband}.

%We show the total optical conductivity 
%$\sigma(\omega)=\sigma^\t{inter}(\omega)+\sigma^\t{intra}(\omega)$
%for various temperatures in Fig.~\ref{fig: sxx inter+intra}.

\subsection{Temperature dependence of Drude weight}
We study the behavior of the Drude weight in the limit $T\to 0$.
The Drude weight is given by the coefficient of $\delta(\omega)$ in Eq.~(\ref{eq: sxx intra}) as
\begin{align}
W_\t{Drude}
&=\frac{1}{6\pi^2}\int dk k^2 \frac{1}{Z(k)^2} \left\{
-(e^{-\beta h}+e^{-\beta \frac U 2})^2 n'_F\left(h_k-\frac U 2\right)
-(e^{\beta h}+e^{-\beta \frac U 2})^2 n'_F\left(h_k+\frac U 2\right)
\right\}
\n
&=\frac{\beta}{6\pi^2} e^{-\beta \frac U 2}\int dk k^2 
\frac{e^{-\beta k}+e^{\beta k}}{(e^{-\beta k}+e^{\beta k}+2e^{-\beta \frac U 2})^2}. 
\end{align}
In the noninteracting case ($U=0$), the Drude weight behaves as $W_\t{Drude} \propto T^2$.
This is obtained from a crude estimation by replacing the factor 
$\frac{e^{-\beta k}+e^{\beta k}}{(e^{-\beta k}+e^{\beta k}+2)^2}$ in the integrand with 1 for $k<T$ and with 0 otherwise.
On the other hand, in the case of strong interactions ($U \to \infinity$),
the Drude weight behaves as $W_\t{Drude} \propto e^{-\beta \frac U 2}T^2$.
Thus the Drude weight is suppressed exponentially
as the interaction $U$ increases.

\subsection{Hall conductivity}
We study the Hall conductivity for a fixed value of $k_z$.
The Hall conductivity has a nonzero contribution from a combination
\begin{align}
\langle T_\tau  b_+ (\tau) \bra + \sigma_x \ket - b_-^\dagger (\tau)
b_-  \bra - \sigma_y \ket + b_+^\dagger \rangle .
\end{align}
Here, we set the momentum transfer as $q=0$ because we focus on the dc Hall conductivity. 
We note that other combinations of current matrices vanish.
After we integrate over the direction $\phi$ of $(k_x,k_y)=k_\parallel (\cos \phi, \sin \phi)$ in current matrices [Eq.~(\ref{eq: matrix elements for sxy})], 
the expectation value is given by
\begin{align}
Q(\tau) &\equiv \frac{2\pi i k_z}{k} \langle b_+ (\tau) b_-^\dagger (\tau)
b_-  b_+^\dagger \rangle +
\langle b_- (\tau) b_+^\dagger (\tau)
b_+  b_-^\dagger \rangle 
\n
&=
\frac{2\pi i k_z}{k} \frac{1}{Z}(
\bra 0 b_-  b_+ (\tau) b_-^\dagger (\tau)
b_-  b_+^\dagger b_-^\dagger \ket 0 e^{\beta h}
+
\bra 0 b_+  b_- (\tau) b_+^\dagger (\tau)
b_+  b_-^\dagger b_+^\dagger \ket 0 e^{-\beta h}
) \n
&=
\frac{2\pi i k_z}{k} \frac{1}{Z}(e^{-2\tau h +\beta h}+ e^{2\tau h -\beta h}).
\end{align}
With the Fourier transformation, we obtain
\begin{align}
Q(i\omega_n)&=
\frac{2\pi i k_z}{k}\frac{1}{Z} 
\int_0^\beta d\tau e^{i\omega_n \tau} \langle b_+ (\tau) b_-^\dagger (\tau)
b_-  b_+^\dagger \rangle +
\langle b_- (\tau) b_+^\dagger (\tau)
b_+  b_-^\dagger \rangle
\n
&=
\frac{2\pi i k_z}{k}\frac{1}{Z} 
\left( \frac{-e^{-\beta h}-e^{\beta h}}{i\omega_n-2h}
+  \frac{-e^{\beta h}-e^{-\beta h}}{i\omega_n+2h}
\right) \n
&=\frac{2\pi i k_z}{k}\frac{1}{Z} 
(e^{\beta h}+e^{-\beta h})\frac{2i\omega_n}{-(i\omega_n)^2+4h^2}
\end{align}
By performing analytic continuation and taking the zero frequency limit,
we obtain the Hall conductivity
\begin{align}
\sigma_{xy} (k_z)&=
\frac{1}{(2\pi)^2}\int k_\parallel dk_\parallel \t{Re}(\frac{Q(\omega)}{-i\omega})\n
&=-\frac{1}{2\pi}
\int k_\parallel dk_\parallel \frac{k_z}{2k^3} 
\frac{e^{\beta h}+e^{-\beta h}}{e^{\beta h}+e^{-\beta h}+2 e^{-\beta \frac U 2}}
\end{align}
where $k_\parallel$ is the radial coordinate for $(k_x,k_y)$ and $k=\sqrt{k_\parallel +k_z^2}$.
In the zero temperature limit,
the Hall conductivity reduces to
\begin{align}
\sigma_{xy} (k_z)&=
-\frac{1}{2\pi}\int_0^\infinity k_\parallel dk_\parallel \frac{k_z}{2k^3} 
%\n
=\left. -\frac{1}{2\pi} \frac{k_z}{2k}\right|_{k_\parallel=0}^{k_\parallel=\infinity} 
%\n
=\frac 1 {4\pi} \t{sgn}(k_z).
\end{align}
If we restore the unit of $e^2/\hbar$,
the Hall conductivity is given by
\begin{align}
\sigma_{xy} (k_z)&=\frac {e^2}{2h} \t{sgn}(k_z),
\end{align}
which remains quantized into $\pm e^2/2h$ in the WMI.

\section{Stability of the Mott gap}
We study the stability of the Mott gap against the interaction
\begin{align}
H_C=\sum_{k,k',q} V(\Vec q) c^\dagger_{\Vec k +\Vec q, \sigma} c^\dagger_{\Vec{k'} -\Vec q, \sigma'}
c_{\Vec{k'} , \sigma'}c_{\Vec k , \sigma}.
\end{align}
We consider the self-energy arising in the second order of this interaction,
\begin{align}
\Sigma (\Vec k,i\omega) &=
\int d^3\Vec{q} \sum_{i\Omega}
V(\Vec q)V(-\Vec q)
G(\Vec k + \Vec q,i\omega+i\Omega)
\Pi(\Vec q, i\Omega),
\end{align}
with the density-density correlation function
\begin{align}
\Pi(\Vec q, i\Omega)&=
d^3\Vec{k'}
\sum_{i\omega'}\t{Tr}[G(\Vec{k'},i\omega') G(\Vec{k'} - \Vec q,i\omega'-i\Omega)].
\end{align}
This is explicitly written as
\begin{align}
\Sigma (\Vec k,i\omega) 
&=
\int d^3\Vec{q} d^3\Vec{k'}\sum_{i\omega',i\Omega}V(\Vec q)V(-\Vec q)
\frac{(i\omega+i\Omega)+(|\Vec k + \Vec q|+\frac U 2)\Vec n (\Vec k + \Vec q) \cdot \Vec \sigma}{(i\omega+i\Omega)^2-(|\Vec k + \Vec q|+\frac U 2)^2}
\frac{1}{(i\omega')^2-(|\Vec {k'}|+\frac U 2)^2}
\n
&\hspace{6em} \times
\frac{1}{(i\omega'-i\Omega)^2-(|\Vec {k'} - \Vec q|+\frac U 2)^2}
\n
&\hspace{6em} \times
2
\left[i\omega'(i\omega'-i\Omega)+ \left(|\Vec {k'}|+\frac U 2 \right) \left(|\Vec {k'} - \Vec q|+\frac U 2 \right)
\Vec n (\Vec {k'}) \cdot \Vec n ( \Vec {k'}- \Vec q) \right]
.
\end{align}
If the instability for the Mott gap were present,
the gap should close at $\Vec k=\Vec 0$ by the consideration from the rotation symmetry.
Therefore, we focus on the self-energy for $\Vec k=\Vec 0$.
 %and the case of contact interaction $V(\bm q)=V$.
By summing over Matsubara frequencies and setting $\Vec k=\Vec 0$, we obtain
\begin{align}
\Sigma (\Vec k=\Vec 0,i\omega) &= 
\int d^3\Vec{q} d^3\Vec{k'}
\frac{|V(\Vec q)|^2}{2} \left(1-\frac{\Vec {k'} \cdot (\Vec {k'} - \Vec q)}{|\Vec {k'}||\Vec {k'} - \Vec q|} \right)
\frac{1}{(i\omega-|\Vec {k'}|-|\Vec {k'} - \Vec q|-U)^2-(|\Vec q|+\frac U 2)^2} \n
&\hspace{3em} \times
\left[i\omega-|\Vec {k'}|-|\Vec {k'} - \Vec q|-U+ \left(|q|+\frac U 2 \right)\Vec{n}_{\Vec{q}} \cdot \Vec \sigma \right].
\end{align}
After performing an integration over $\Vec k'$, the terms $|\Vec {k'} - \Vec q|$ and $\Vec {k'} \cdot (\Vec {k'} - \Vec q)$ no longer have a dependence on the angle of $\Vec q$, because they only depend on the relative angle between $\Vec k'$ and $\Vec q$.
Then the only term depending on the angle of $\Vec q$ after the $\Vec k'$ integration is $\Vec{n}_{\Vec{q}} \cdot \Vec \sigma$,
which vanishes upon the integration over the angle of $\Vec q$.
Thus the self-energy $\Sigma (\Vec k=\Vec 0,i\omega)$ is diagonal with respect to the spin degrees of freedom.
Furthermore, the imaginary part of $\Sigma (\Vec k,\omega)$ (after the analytic continuation) appears only at
$\omega=|\Vec {k'}|+|\Vec {k'} - \Vec q|+|\Vec q|+ \frac {3U} 2 \ge \frac {3U} 2$
and $\omega=|\Vec {k'}|+|\Vec {k'} - \Vec q|-|\Vec q|+ \frac {U} 2 \ge \frac {U} 2$;
The imaginary part of $\Sigma (\Vec k=\Vec 0,\omega)$ is zero for $\omega <\frac {U} 2$.
Therefore, the gap of $\frac U 2$ in the Green's function is stable against the inclusion of the interaction $H_C$.

We note that the the perturbation theory with respect to $V(\Vec q)$ is valid
because of the absence of the infrared divergence.
In the case of the contact quartic interaction $V(\Vec q)=V$,
we notice that the infrared divergence does not appear for $i\omega=0$
because of the gap of $\frac U 2$ in the energy denominator.
In the case of the repulsive Coulomb interaction $V(\Vec q)=\frac{4\pi e^2}{q^2}$, the infrared divergence is also absent,
because the density-density correlation function behaves 
$\Pi(\Vec q, i\Omega) \propto q^2$ for small $q$ and $\Omega$, and 
the integral is convergent around $q=0$.
%The absence of the infrared divergence validates .
%Therefore, the gap of $\frac U 2$ in the Green's function is stable against the inclusion of the interaction $H_C$.

%\bibliography{TI}

\end{document}